\documentclass[12pt]{article}
\pdfoutput=1
\usepackage[centertags]{amsmath}
\usepackage[square, comma, sort&compress,numbers]{natbib}
\usepackage{array,multirow}
\numberwithin{equation}{section}
\usepackage{amssymb,amsmath,amsthm}
\usepackage{graphicx}
\usepackage{color}
\usepackage[normalem]{ulem}

\usepackage{epsfig}
\usepackage{bbold}

\usepackage{wrapfig}
\usepackage{float}
\usepackage{soul}

\usepackage{tikz,pgf}
\usetikzlibrary{shapes}
\usetikzlibrary{calc}
\usetikzlibrary{decorations.pathmorphing}
\usetikzlibrary{decorations.pathreplacing,shapes.misc}
\usetikzlibrary{positioning}
\usetikzlibrary{arrows}
\usetikzlibrary{decorations.markings}
\usetikzlibrary{shadings}

\usetikzlibrary{intersections}

\def\be{\begin{equation}}
\def\ee{\end{equation}}
\def\ba{\begin{array}}
\def\ea{\end{array}}

\def\dps{\displaystyle}

\def\sd{D^\dagger}

\newcommand{\brst}{\mathsf{\Omega}}

\def\half{\frac{1}{2}}

\newcommand{\fo}{\theta}
\newcommand{\ao}{\Gamma}
\newcommand{\Dirac}{\widehat{D}}

\usepackage{jheppub}
\usepackage{amsmath,amsfonts,amssymb,amsthm}
\newcommand{\bref}[1]{\textbf{\ref{#1}}}

\newcommand{\dl}[1]{\frac{\dd}{\dd #1}}

\newcommand{\dd}{\partial}
\renewcommand{\d}{\partial}
\newcommand{\gh}[1]{\mathrm{gh}(#1)}

\newcommand{\inner}[2]{\langle #1{,}\,#2\rangle}

\def\cE{\mathcal{E}}
\def\cF{\mathcal{F}}
\def\cG{\mathcal{G}}
\def\cH{\mathcal{H}}

\def\cP{\mathcal{P}}

\def\cS{\mathcal{S}}

\def\bC{\mathbb{C}}

\usepackage{mathtools}	
\usepackage{bm}

\newcommand*\dif{\mathop{}\!\mathrm{d}}

\DeclareMathOperator{\Ker}{Ker}
\DeclareMathOperator{\Ima}{Im}

\newcommand{\Liealg}{\mathfrak}

\makeatletter
\def\@fpheader{\vspace{-.1cm}}
\makeatother

\makeatletter
\def\widebreve{\mathpalette\wide@breve}
\def\wide@breve#1#2{\sbox\z@{$#1#2$}%
     \vbox{\m@th\ialign{##\crcr
\kern0.08em\brevefill#1{0.8\wd\z@}\crcr\noalign{\nointerlineskip}%
                    $\hss#1#2\hss$\crcr}}}
\def\brevefill#1#2{$\m@th\sbox\tw@{$#1($}%
  \hss\resizebox{#2}{\wd\tw@}{\rotatebox[origin=c]{90}{\upshape(}}\hss$}
\makeatletter

\makeatletter
\def\widebar{\mathpalette\wide@bar}
\def\wide@bar#1#2{\sbox\z@{$#1#2$}%
     \vbox{\m@th\ialign{##\crcr
\kern0.14em\barfill#1{0.88\wd\z@}\crcr\noalign{\nointerlineskip}%
                    $\hss#1#2\hss$\crcr}}}
\def\barfill#1#2{$\m@th\sbox\tw@{$#1($}%
  \hss\resizebox{#2}{\wd\tw@}{\rotatebox[origin=c]{0}{\upshape-}}\hss$}
\makeatletter

\usepackage{mleftright}

\title{Lagrangian BRST formulation of massive higher spin fields of generic symmetry type}

\author[a,b]{Alexander \ Chekmenev}

\affiliation[a]{I.E. Tamm Department of Theoretical Physics, \\P.N. Lebedev Physical
  Institute,\\ Leninsky ave. 53, 119991 Moscow, Russia}
\affiliation[b]{Department of Higher Mathematics, \\
  Moscow Institute of Physics and Technology, \\
  Institutskiy per. 7, Dolgoprudnyi, \\141700 Moscow region, Russia}
\emailAdd{chekmenev@phystech.edu}

\abstract{We describe the procedure of dimensional reduction of massless fields in $(D+1)$ dimensional Minkowski space to massive ones in $D$ dimensions in the first-quantized setting. The procedure is compatible with Lagrangian and in a straightforward way determines the inner product for massive fields. Howe duality and the extensive usage of BRST formalism allows to keep the description concise. 
We consider both bosonic and fermionic mixed-symmetry fields.}
  
\begin{document}

\maketitle

\section{Introduction}


The goal of this paper is to give a systematic description of the procedure of dimensional reduction from $\mathbb R^{D,1}$ to $\mathbb R^{D-1,1}$ for both bosonic and fermionic mixed symmetry fields
\cite{Aragone:1988yx,Pashnev1989}. See \cite{Bekaert:2003uc} for a review.

A massive field propagating in flat space is associated with a massive representation of the Poincare algebra \cite{Wigner:1939cj, Bargmann:1948ck} (for review, see \cite{Bekaert:2006py}). A Lagrangian formulation for symmetric massive spin fields was first obtained by Singh and Hagen \cite{Singh:1974qz, Singh:1974rc}. Studying the corresponding massless limit Fronsdal obtained covariant Lagrangians for massless fields of any integer spin in 4 dimensions \cite{Fronsdal:1978rb}.

As for massless mixed symmetry fields, although the existence of corresponding irreducible representations in $d > 5$ was known since Wigner \cite{Wigner:1939cj}, the covariant and gauge invariant field equations realizing such representations on local fields have been proposed much later \cite{Labastida:1986ft, Labastida:1989kw}. The rigorous proof that such fields indeed describe respective representations of the Wigner little group for general spins was given in \cite{Bekaert:2002dt} (see also \cite{Alkalaev:2008gi}).


Making use of the BRST formalism allows one to find a theory that admits a standard Lagrangian of the form $\langle \Phi, \brst \Phi \rangle$. This has the same structure as the analogous Lagrangian for Fronsdal higher spin fields proposed in \cite{Bengtsson:1986ys,Ouvry:1986dv,Henneaux:1987cpbis}. The BRST operator $\brst$ entering the action can be identified with an appropriate truncation of the open bosonic string BRST operator in the tensionless limit \cite{Sagnotti:2003qa,Bonelli:2003kh}. See \cite{Alkalaev:2008gi} for the proof that this Lagrangian indeed describes an irreducible mixed-symmetry field provided the appropriate set of algebraic constraints are imposed on $\Phi$.
Lagrangian formulation of this type for fermionic fields was proposed more recently\footnote{Lagrangians for higher spin fermionic fields were first constructed in \cite{Siegel:1986de}.} \cite{Francia:2002pt,Sagnotti:2003qa,Reshetnyak:2012ec,Alkalaev:2018bqe}. 
For the previous results on free higher spin massless fields propagating in Minkowski and (A)dS backgrounds within various formulations, see e.g. 
\cite{Lopatin:1988hz,Buchbinder:2001bs,Zinoviev:2001dt,Alkalaev:2003qv,Skvortsov:2008sh,Campoleoni:2008jq,Skvortsov:2009zu,Boulanger2009,Bekaert:2015fwa,Joung:2016naf, Fang:1978wz,Vasiliev:1987tk,Metsaev:1998xg,Alkalaev:2001mx,Alkalaev:2003qv,Buchbinder:2004gp,Buchbinder:2007vq,Moshin:2007jt,Skvortsov:2008vs,Zinoviev:2009vy,Campoleoni:2009gs,Skvortsov:2010nh,Reshetnyak:2012ec,Reshetnyak:2018fvd,Reshetnyak:2018yhz,Najafizadeh:2018cpu}.



As for massive fields, each corresponding representation in $D > 4$ dimensions may be obtained as the Kaluza–Klein mode in a dimensional reduction from $D+1$ down to $D$ dimensions \cite{Bekaert:2006py}. There is no loss of generality because the massive little algebra $\Liealg{so}(D-1)$ in $D$ dimension is identified with the massless $(D+1)$-dimensional little algebra. Such a Kaluza–Klein-like mechanism leads to a Stuckelberg formulation of a massive field.

In this paper we describe the procedure of dimensional reduction of massless fields in $(D+1)$ dimensional Minkowski space to massive ones in $D$ dimensions in the first-quantized setting\footnote{Analogous procedure in the anti-de Sitter space was discussed in \cite{Alkalaev:2011zv} (see also \cite{Barnich:2006pc}).}. Howe duality and the extensive usage of BRST formalism allows to keep the description concise.
We consider both bosons and fermions\footnote{The BRST description of bosonic massive fields in similar setting was made in \cite{Chekmenev:thesis, Metsaev:2012uy}.}. We calculate the eigenvalues of quartic Casimir operators.
The inner product for massless bosonic case is known since \cite{Bengtsson:1986ys}. Its natural generalization to fermionic case was found recently in \cite{Alkalaev:2018bqe}. 
The dimensional reduction is compatible with Lagrangian and in a straightforward way determines the inner product for massive fields.
See e.g. \cite{Zinoviev:2009vy,Zinoviev:2008ve, Metsaev:2017cuz, Zinoviev:2001dt, Ponomarev:2010st, Alkalaev:2011zv, Pashnev:1997rm, Buchbinder:2005ua,Buchbinder:2006ge,Buchbinder:2006nu} for earlier investigations of massive higher spin fields formulations on different backgrounds.

The paper is structured as follows.
There are two main parts devoted to bosonic and fermionic fields.
The exposition follow similar lines.

In Section \ref{sec:b_Algebraic_preliminaries} we describe $\Liealg{o}(d - 1,1)-\Liealg{sp}(2n)$ bimodule (which is also a Poincar\'e one) on the functions of auxiliary (anti)commuting variables, which serves as a representation space of the constrained system described later.
In Section \ref{sec:b_Massless_fields} we formulate the constraint system describing massless fields. We describe the triplet and metric-like formulations which are used to get Lagrangian formulations. For bosonic fields there is a natural inner product for the triplet formulation, but for fermionic fields a natural inner product is for the metric-like formulation.
In Section \ref{sec:b_Dimensional_reduction} we perform the dimensional reduction, giving the Lagrangian gauge invariant equations of mixed-symmetry type.

In appendices \ref{app:comm} and \ref{app:Casimir} are given explicitly commutation relations and expressions for Casimir operators of symplectic (super)algebras. Appendix \ref{app:Dimensional_reduction} contain details of dimensional reduction.


\section{Bosonic fields}

We perform the first-quantized description of massless bosonic fields of mixed symmetry. Then we employ the BRST description and perform homological reduction to obtain other formulations. One of the formulations allows natural Lagrangian description.
Then we perform the dimensional reduction to obtain the description of massive fields which replicates almost all constructions of the massless case.

We consider the mostly plus signature of the metric tensor in Minkowski space: $\eta = (- + \ldots +)$.

\subsection{Algebraic preliminaries} \label{sec:b_Algebraic_preliminaries}

\subsubsection{Tensor fields}

Let us introduce a convenient notation which allows to pack tensor indices.

Let us introduce Grassmann even variables $a^a_I$ and $\bar a^J_b$, where $a,b=0, ..., d-1$, $\;I,J=0, ...,n$ satisfying the canonical commutation relations
\begin{equation}
	[{\bar a}_a^I, a^b_{J}] = \delta^I_J\,\delta_a^b\,.
\end{equation}

In the following we will raise and lower lowercase indices with the Minkowski tensor $\eta_{ab} = (- + \ldots +)$, e.g. $a_I{}_a \equiv \eta_{ab} a^b_I$.

Let us denote the space of polynomials in $a_I^a$ as $\cP^d_n(a_I)$.
The component decomposition of a general element $\phi(a) \in \cP^d_n(a)$ have the form
\begin{equation}
  \phi(a)= \sum_{m_0 = 0}^{M_0} \ldots \sum_{m_n = 0}^{M_n} \phi{}_{a_1\;\ldots\; a_{m_0};\;\ldots\ldots\;;\, b_1 \;\ldots\; b_{m_{n}}} a^{a_1}_0 \ldots a^{a_{m_0}}_0 
  \;\ldots\;
  a^{b_1}_{n} \ldots a^{b_{m_{n}}}_{n}\;,
\end{equation}
where $M_0, \ldots M_n \in \mathbb N$.

The associative algebra generated by $a^a_I$, $\bar a^J_b$ can be represented on $\cP^d_{n}(a_I)$ in a natural way if one defines the action of the generators according to
\begin{equation} \label{b_Generators_action}
	a_I^a \phi(a)\coloneqq a_I^a \phi(a)\,,\qquad
  \bar a^I_a \phi(a) \coloneqq \frac{\partial}{\partial a_I^a} \phi(a)\,.
\end{equation}

\subsubsection{Lorentz algebra and symplectic algebra}

The Lorentz algebra $\Liealg{so}(d-1,1)$ can be embedded as a Lie subalgebra into the above associative algebra by postulating its basis elements to be
\begin{equation}
  M_{ab} = a_I{}_a \bar a^{I}_b - a_I{}_b \bar a^{I}_a\,.
\end{equation}
This also defines the representation of $\Liealg{so}(d-1,1)$ on $\cP^d_{n}(a_I)$ via \eqref{b_Generators_action}.

Simultaneously, the symplectic algebra  $\Liealg{sp}(2n+2)$ can also be embedded into the associative algebra, and, hence, is also represented on $\cP^d_{n}(a_I)$. The basis elements are given by
\begin{equation}
	T_{IJ} = a_I^a a_{Ja}\,,\qquad
  T_I{}^J = \frac12 \left(a^a_I \bar a^J_a + \bar a^J_a a^a_I\right),\qquad
  T^{IJ} = \bar a^I_a \bar a^{Ja}\,,
\end{equation}
with the commutation relations given in Appendix \bref{app:comm}.
The two algebras mutually commute forming a reductive dual pair~\cite{Howe,Howe1}.
So the space $\cP^d_{n}(a_I)$ is now $\Liealg{so}(d-1,1)-\Liealg{sp}(2n+2)$ bimodule. 

\subsubsection{Poincar\'e algebra}

The Poincar\'e algebra $\Liealg{iso}(d-1,1)$ can be realized on the same set of auxiliary variables. To this end, we split the original  variables as $a^a_0\equiv x^a,\;a^a_I\equiv a^a_i\,,\,I>0$ with $i=1,...,n$. Then, translations and Lorentz rotations are given by 
\begin{equation}
  P_a = \partial_a\;, \quad\qquad 
  M_{ab}
  = x_a \partial_b - x_b \partial_a
  + a_i{}_a \bar a^{i}_b - a_i{}_b \bar a^{i}_a\,,
\end{equation}
and naturally act in the space $\cP_n^d(x,a)$ of smooth functions in $x^a$ with values in $\cP_n^d(a_i)$. 

We also introduce special notation for some of $\Liealg{sp}(2n+2)$ basis elements 
\begin{equation} \label{b_sp_basis_notation}
\begin{gathered}
  \Box \equiv T^{00} = \partial_a \partial^a\,,\qquad
  D^i \equiv T^{0i} = \bar a_i^a\partial_a\,, \qquad
  \sd_i\equiv T_i{}^0 = a_i^a \partial_a\,,\qquad \\
  N_i{}^j\equiv T_i{}^j = a_i^a \bar a_{ja} \qquad i\neq j\,,\qquad
  N_i \equiv T_i{}^i - \frac{d}{2} = a_i^a \bar a_{ia}\,.
\end{gathered}
\end{equation}

\subsection{Massless fields} \label{sec:b_Massless_fields}

It is well-known that massless bosonic fields of arbitrary symmetry type can be concisely described in the first-quantized way as a set of tensor fields $\phi \in \cP_n^d(x,a)$ subjected to constraints which come from Howe duality between symplectic and orthogonal algebras.

\paragraph{Differential constraints} are the Klein-Gordon equation and divergence-free condition
\begin{equation} \label{b_diff_constraints}
	\Box \phi = 0\,,\qquad
	D^i \phi = 0\,.
\end{equation}
\paragraph{Algebraic constraints} impose that fields are traceless tensor fields of given ranks which satisfy Young symmetry conditions:
\begin{equation} \label{b_alg_constraints}
  T^{ij} \phi = 0\,,\qquad
	N_i \phi = s_i \phi\,,\qquad
  N_i{}^j \phi = 0 \quad (i < j)\,.
\end{equation}
\paragraph{Gauge invariance.}
We impose the last constraint in a dual way via the gauge transformations which read
\begin{equation}
  \delta \phi = D^\dag_i \chi^i\,, \qquad i = 1, \ldots, n\,,
\end{equation}
where the gauge parameters $\chi^i$ satisfy the same constraints as the fields $\phi$ except for spin weight and Young symmetry constraints which are replaced by
\begin{gather}
  N_i \chi^j = (s_i - \delta_i^j) \chi^j\,,\qquad
  N_i{}^j \chi^k = - \delta^k_i \chi^j \quad (i < j)\,.
\end{gather}

\subsubsection{Simplest BRST formulation}

Let us introduce the Grassmann odd variables $b_i$ of ghost number $\gh{b_i} = -1$. The gauge symetries can be realised via the BRST operator
\begin{equation}
	Q = D^\dag_i \frac{\partial}{\partial b_i}\,.
\end{equation}
It acts on a subspace of $\cP_n^d(x, a) \otimes \mathbb R[b_i] \ni \Phi(x, a \,|\, b)$, where $\mathbb R[b_i]$ is the Grassmann algebra generated by $b_1, \ldots, b_n$, where functions $\Phi$ are subjected to BRST invariant extension of constraints \eqref{b_diff_constraints}, \eqref{b_alg_constraints} which remain the same except for the Young symmetry and spin conditions which receive ghost extensions. Namely
\begin{equation}
	\mathcal N_i \Phi = s_i \Psi\,,\qquad
  \mathcal N_i{}^j \Phi = 0 \quad (i < j)\,,
\end{equation}
where
\begin{equation}
	\mathcal N_i{}^j = N_i{}^j + b_i \frac{\partial}{\partial b_j}\,,\qquad
  \mathcal N_i = N_i + b_i \frac{\partial}{\partial b_i}\,.
\end{equation}

This formulation provides an explicit description of gauge for gauge transformations as seen from the following.
The space of field configurations becomes $\mathbb Z$-graded. Arbitrary vector $\Phi$ decomposes as
\begin{equation}
	\Phi = \sum_{k = 0}^{n} \Phi^{(-k)}\,,\qquad \gh{\Phi^{(-k)}} = -k\,.
\end{equation}
Here and what follows we denote by superscript in parenthesis the definite ghost degree component of a field. We interpret zero ghost degree components as physical fields, and negative components as gauge (for gauge) parameters. So the equations of motion are
\begin{equation}
	Q \Phi^{(0)} = 0
\end{equation}
and the gauge transformations are
\begin{equation}
	\delta \Phi^{(0)} = Q \Phi^{(-1)}\,.
\end{equation}

\subsubsection{Triplet formulation}

It is useful to impose all differential constraints by means of BRST operator, while all algebraic constraints we impose on the representation space. This is known as the triplet formulation \cite{Alkalaev:2008gi,Bengtsson:1986ys,Francia:2002pt,Sagnotti:2003qa}.

Let us introduce new anticommuting ghost variables $c_0, c_i$, $i = 1, ..., n$ with ghost numbers $\gh{c_0} = \gh{c_i} = 1$.

Consider the BRST operator
\begin{equation} \label{b_BRST}
	\brst = c_0 \Box + c_i D^i + D^\dag_i \frac{\partial}{\partial b_i} - c_i \frac{\partial}{\partial b_i} \frac{\partial}{\partial c_0}\,,
\end{equation}
which acts on a subspace of $\cP_n^d(x, a) \otimes \mathbb R[b_i,c_0,c_i] \ni \Psi(x, a \,|\, b_i, c_0, c_i)$ such that
\begin{equation} \label{b_Triplet_algebraic}
  \widetilde N_i \Psi = s_i \Psi\,, \qquad
  \widetilde N_i{}^j \Psi = 0 \quad (i < j)\,, \qquad
  \widetilde T^{ij} \Psi = 0\,,
\end{equation}
where
\begin{equation}
  \widetilde N_i{}^j = N_i{}^j + b_i \frac{\partial}{\partial b_j} + c_i \frac{\partial}{\partial c_j}\;,\qquad
  \widetilde N_i = \widetilde N_i{}^i\;,\qquad
  \widetilde T^{ij} = T^{ij} + \frac{\partial}{\partial c_i} \frac{\partial}{\partial b_j} + \frac{\partial}{\partial c_j} \frac{\partial}{\partial b_i}
\end{equation}
are BRST invariant extensions of algebraic constraints \eqref{b_alg_constraints}.

So the equations of motion have the form
\begin{equation} \label{b_Triplet_equation}
	\brst \Psi^{(0)} = 0\,.
\end{equation}
By construction it is invariant with respect to gauge transformations
\begin{equation}
	\delta \Psi^{(0)} = \brst \Psi^{(-1)}\,.
\end{equation}

\subsubsection{Lagrangian formulation} \label{sec:b_Lagrangian}

Another useful property of the description by means of the BRST operator \eqref{b_BRST} is that the space it acts on admits a natural nondegenerate inner product.


The BRST operator \eqref{b_BRST} is formally self-adjoint with respect to the nondegenerate inner product
\begin{equation}
  \langle \psi, \varphi \rangle
  = \int \dif^d x \int \dif c_0 \langle \psi, \varphi \rangle'\,,
\end{equation}
where $\langle *, * \rangle'$ is the inner product on the Fock module generated by $a_i^a, c_i, b_i$ from the ''vacuum vector'' $|0\rangle$. Conjugation rules are
\begin{equation}
  (a_i^a)^\dag = \frac{\partial}{\partial a_i^a}\,, \qquad
  (c_i)^\dag = - \frac{\partial}{\partial b_i}\,, \qquad
  (b_i)^\dag = - \frac{\partial}{\partial c_i}\,
\end{equation}
and (anti)commutation identities completely determine the inner product.
The formal symmetry of the BRST operator \eqref{b_BRST} with respect to it is straightforward.

So the equations of motion $\brst \Phi^{(0)} = 0$ follow from the action
\begin{equation}
  S = \frac12 \langle \Phi^{(0)}, \brst \Phi^{(0)} \rangle\,.
\end{equation}
This action is invariant under the gauge transformations $\delta \Phi^{(0)} = \brst \Phi^{(-1)}$.
If one takes into account the off-shell algebraic constraints \eqref{b_Triplet_algebraic} it is equivalent to the Fang-Fronsdal-Labastida action.

\subsubsection{Homological reduction}

One of features of BRST formulation is the ability to obtain other dynamically equivalent formulations by means of homological algebra. In particular, we will use it to derive a Lagrangian formulation for fermionic fields. Here we employ the method of homological reduction \cite{Barnich:2004cr}.

In short, homological reduction method is as follows. Let $(\cH, \brst)$ be the BRST complex with linear BRST operator $\brst$ acting on the space $\cH$, which is a $\mathbb Z$-graded space according to the ghost number. And let $\cH$ be split into three subspaces: $\cH = \cE \oplus \cF \oplus \cG$ in such a way that a linear operator $\overset{\cG \cF}{\brst}: \cF \to \cG: f \mapsto \mathrm{pr}_\cG \brst f$ for $f \in \cF$, is invertible. It turns out that fields associated with $\cF$ and $\cG$ are generalized auxiliary fields that can be eliminated resulting in dynamically equivalent formulation $(\cE, \widetilde \brst)$ of the same theory. Explicit form of the reduced BRST operator is
\begin{equation}
	\widetilde \brst = \overset{\cE \cE}{\brst} - \overset{\cE \cF}{\brst} (\overset{\cG \cF}{\brst})^{-1} \overset{\cG \cE}{\brst},
\end{equation}
where $\overset{\cE \cE}{\brst}$ and $\overset{\cG \cE}{\brst}$ defined in the same way as $\overset{\cG \cF}{\brst}$ above.


It is convenient to define that decomposition of $\cH$ by a certain piece of $\brst$. Namely, suppose that $\cH$ admits an additional $\mathbb Z$-grading
\begin{equation}
	\cH = \bigoplus_{i = -N}^\infty \cH_i \qquad N \in \mathbb N,
\end{equation}
such that $\brst$ decomposes into homogeneous components as follows
\begin{equation}
	\brst = \brst_{-1} + \brst_{0} + \brst_1 + \ldots.
\end{equation}
Then the lowest degree part $\brst_{-1}$ is nilpotent and defines the triple decomposition
\begin{equation}
	\cE \oplus \cG = \Ker \brst_{-1},\qquad
  \cG = \Ima \brst_{-1},\qquad
  \cE \cong \Ker \brst_{-1} / \Ima \brst_{-1} \equiv H(\brst_{-1}).
\end{equation}
Subspaces $\cE \oplus \cG \subset \cH$ and $\cG \subset \cH$ defined canonically, there is an ambiguity of embedding of $\cE$ and $\cF$ into $\cH$.

If one is interested in local gauge field theories (which is usually the case) one requires that generalized auxiliary fields can be eliminated algebraically. In case at hand that means that $\brst_{-1}$ does not involve space-time derivatives.

\subsubsection{Metric-like formulation} \label{sec:b_Metric-like}

Let us take as additional degree the homogeneity in $c_0$. The BRST operator \eqref{b_BRST} decomposes as
\begin{equation}
	\brst_{-1}
= - c_i \frac{\partial}{\partial b_i} \frac{\partial}{\partial c_0}\,,\qquad
  \brst_0
= c_i D^i
+ D^\dag_i \frac{\partial}{\partial b_i}\,,\qquad
  \brst_1
= c_0 \Box\,.
\end{equation}

Performing the homological reduction technique we can reduce the description to the subspace $\cE = H(\brst_{-1})$.
See \cite{Alkalaev:2008gi} for useful details.
Let us perform some component analysis.

Let us decompose ghost degree $0$ and $-1$ components as 
\begin{equation}
	\Psi^{(0)} = \varphi + \ldots\,,\qquad
  \Psi^{(-1)} = b_i \varepsilon^i + \ldots\,,
\end{equation}
where ellipses denote ghost free components.
From \eqref{b_Triplet_algebraic} it follows that $\varphi$ and $\varepsilon^i$ satisfy modified trace conditions
\begin{equation}
	T^{(ij} T^{kl)} \varphi = 0\,,\qquad
  T^{(ij} \varepsilon^{k)} = 0
\end{equation}
and Young symmetry and spin weight conditions
\begin{equation}
	N_i{}^j \varphi = 0 \quad (i < j)\,,\qquad
  N_i \varphi = s_i \varphi\,,
\end{equation}
\begin{equation}
	N_i{}^j \varepsilon^k + \delta_i^k \varepsilon^j = 0 \quad (i < j)\,,\qquad
  N_i \varepsilon^j = (s_i - \delta_i^j) \varepsilon^j\,.
\end{equation}

From \eqref{b_Triplet_equation} it follows that
\begin{equation}
	\left( \Box - D^\dag_i D^i + \frac12 D^\dag_i D^\dag_j T^{ij} \right) \varphi = 0\,,
\end{equation}
which is invariant with respect to the gauge transformations
\begin{equation}
	\delta \varphi = D^\dag_i \varepsilon^i\,.
\end{equation}
This reproduces the Labastida formulation \cite{Labastida:1989kw}.

\subsection{Massive fields} \label{sec:b_Dimensional_reduction}

The trick to describe a massive field in $d$-dimensional space is to start with the system describing a massless field in $(d+1)$ dimensions, to fix some spacelike direction $V^A$, $A = 0, \ldots, d$, and to impose an additional constraint
\begin{equation} \label{Momentum_fix_covariant}
	\left( V^A \frac{\partial}{\partial X^A} - m \right) \phi = 0\,.
\end{equation}
It commutes with all the constraints above so this addition is consistent.

It's convenient to choose $V^A = \delta^{A}_d$. The constraint \eqref{Momentum_fix_covariant} above becomes
\begin{equation} \label{Momentum_fix}
	\left( \frac{\partial}{\partial x^{d}} - m \right) \phi = 0\,.
\end{equation}

Technically we are going to repeat the previous subsection \ref{sec:b_Massless_fields} with the additional constraint \eqref{Momentum_fix}. The only subtlety is that we reduce the Poincar\'e algebra $\Liealg{iso}(d,1)$ to its subalgebra $\Liealg{iso}(d-1,1)$. To make this explicit we denote oscillators ''along the distinguished direction'' as $a^{d}_i \equiv z_i$ and replace $\d_{d}$ with $m$. Note that such replacement makes the $(D+1)$-dimensional BRST operator formally not self-adjoint. But this is fixed by changing the conjugation rule for $z_i$.

So effectively we have a $d$-dimensional system of field configurations subjected to the following differential and algebraic constraints
\begin{equation} \label{b_massive_EOM_diff}
	\left( \Box + m^2 \right) \phi = 0\,,\qquad
	\left( D^i + m \frac{\partial}{\partial z_i} \right) \phi = 0\,,
\end{equation}
\begin{equation} \label{b_massive_EOM_alg}
  \left( T^{i j} + \frac{\partial}{\partial z_i} \frac{\partial}{\partial z_j} \right) \phi = 0\,,\quad
	\left( N_i + Z_i \right) \phi = s_i \phi\,,\quad
  \left( N_i{}^j + z_i \frac{\partial}{\partial z_j} \right) \phi = 0 \;\; (i < j)\,,
\end{equation}
(here $Z_i \equiv z_i \frac{\partial}{\partial z_i}$)
along with the gauge equivalence
\begin{equation} \label{b_massive_gauge}
  \delta \phi = \left( D^\dag_i + m z_i \right) \chi^i\,, \qquad i = 1, \ldots, n\,,
\end{equation}
where the gauge parameters $\chi^i$ satisfy the same constraints as the fields $\phi$ except for the spin weight and Young symmetry constraints which are replaced by
\begin{equation} \label{b_massive_EOM_gauge}
  \left( N_i + Z_i \right) \chi^j = (s_i - \delta_i^j) \chi^j\,,\qquad
  \left( N_i{}^j + z_i \frac{\partial}{\partial z_j} \right) \chi^k = - \delta^k_i \chi^j \quad (i < j)\,.
\end{equation}

\subsubsection{Evaluation of Casimir operators}

Using explicit expressions \eqref{Casimir2_regular} and \eqref{Casimir4_regular} for the quadratic and quartic Casimir operators in terms of the $\Liealg{sp}(2n)$ basis elements and taking into account constraints \eqref{b_massive_EOM_diff}, \eqref{b_massive_EOM_alg} a direct calculation yields
\begin{equation}
	C_2 \phi = - m^2 \phi\,,
\end{equation}
\begin{equation} \label{b_massive_C4}
	C_4 \phi =
- m^2 \sum_i s_i (s_i + d - 1 - 2 i) \phi
+ (D^\dag_i + m z_i) \mleft(m (d + 2 s_1 - 4) \alpha^i - \beta^i + 2 m \gamma^i\mright),
\end{equation}
where
\begin{equation}
  \alpha^i \!=\! \frac{\partial}{\partial z_i} \phi\,,\quad
  \beta^i \!=\! (D^\dag_j + m z_j) \frac{\partial}{\partial z_j} \alpha^i,\quad
	\gamma^i
\!=\! \sum_{j > i} (N_j{}^i + z_j \frac{\partial}{\partial z_i}) \alpha^j
- (s_1 - s_i + i - 1) \alpha^i.
\end{equation}
It is straightforward to check that $\alpha^i$ (as well as $\beta^i$ and $\gamma^i$) satisfy \eqref{b_massive_EOM_gauge} and all other constraints on gauge parameters provided $\phi$ satisfies \eqref{b_massive_EOM_diff}, \eqref{b_massive_EOM_alg} (this is obvious for $\alpha^i$; use the $\Liealg{sp}(2n)$ algebra relations to check the others). It follows that the last term in \eqref{b_massive_C4} is pure gauge. So the quartic Casimir operator acts diagonally on the quotient space. As it should be, because $C_4$ commutes with all the constraints (due to Howe duality), so it leaves the subspace of pure gauge fields invariant. Therefore its action is well defined on the quotient space, which is the space of irreducible representation of the $\Liealg{iso}(d - 1, 1)$ algebra.

Thereby we calculated the eigenvalue of $C_4$ understood as acting on the space of equivalence classes $\overline \phi$ of field configurations modulo gauge transformations\footnote{
Analogous technique of evaluation of eigenvalues of Casimir operators was performed in \cite{Alkalaev:2009vm,Alkalaev:2011zv,Alkalaev:2017hvj}.}:
\begin{equation}
	C_4 \overline \phi =
- m^2 \sum_{i = 1}^n s_i (s_i + d - 1 - 2 i) \overline \phi\,.
\end{equation}
See also section \bref{sec:b_massive_Casimir_indirect} for a more technically transparent but indirect argument.

\subsubsection{Triplet formulation}

The BRST operator associated to the constrained system \eqref{b_massive_EOM_diff}, \eqref{b_massive_EOM_alg}, \eqref{b_massive_EOM_gauge} is
\begin{equation} \label{b_BRST_massive}
  \brst_m
= c_0 \left( \Box + m^2 \right)
+ c_i \left( D^i + m \frac{\partial}{\partial z_i} \right)
+ \left( D^\dag_i + m z_i \right) \frac{\partial}{\partial b_i}
- c_i \frac{\partial}{\partial b_i} \frac{\partial}{\partial c_0}\,.
\end{equation}
Which is a massive counterpart of \eqref{b_BRST}.

It acts on a subspace of $\cP_n^d(x, a) \otimes \mathbb R[b_i,c_0,c_i] \ni \Psi(x, a \,|\, b_i, c_0, c_i)$ singled out by the BRST invariant extensions of algebraic constraints \eqref{b_massive_EOM_alg}
\begin{equation} \label{b_massive_Triplet_algebraic}
  \widetilde N_i \Psi = s_i \Psi\,, \qquad
  \widetilde N_i{}^j \Psi = 0 \quad (i < j)\,, \qquad
  \widetilde T^{ij} \Psi = 0\,,
\end{equation}
where
\begin{equation}
\begin{gathered}
  \widetilde N_i{}^j = N_i{}^j + z_i \frac{\partial}{\partial z_j} + b_i \frac{\partial}{\partial b_j} + c_i \frac{\partial}{\partial c_j}\;,\qquad
  \widetilde N_i = \widetilde N_i{}^i\;,\\
  \widetilde T^{ij} = T^{ij} + \frac{\partial}{\partial z_i} \frac{\partial}{\partial z_j} + \frac{\partial}{\partial c_i} \frac{\partial}{\partial b_j} + \frac{\partial}{\partial c_j} \frac{\partial}{\partial b_i}\,.
\end{gathered}
\end{equation}

\subsubsection{Lagrangian formulation}

We follow similar lines as in \ref{sec:b_Lagrangian}. We consider $z_i$ and $\frac{\partial}{\partial z_i}$ as another canonically conjugated pair of creation and annihilation operators in the Fock space. The conjugation rule
\begin{equation}
	(z_i)^\dag = - \frac{\partial}{\partial z_i}
\end{equation}
determines the inner product $\langle *, * \rangle'$ on the Fock space.

%

The inner product on the whole space is taken to be
\begin{equation} \label{b_Inner_massive}
  \langle \psi, \varphi \rangle
  = \int \dif^d x \int \dif c_0 \langle \psi, \varphi \rangle'\,.
\end{equation}
It's straightforward to see that the BRST operator \eqref{b_BRST_massive} is formally symmetric with respect to the inner product \eqref{b_Inner_massive}.

So the equations of motion $\brst_m \Phi^{(0)} = 0$ follows from the action
\begin{equation}
  S = \frac12 \langle \Phi^{(0)}, \brst_m \Phi^{(0)} \rangle\,,
\end{equation}
which is invariant under gauge transformations $\delta \Phi^{(0)} = \brst_m \Phi^{(-1)}$.

\subsubsection{Gauge fixing} \label{sec:b_Gauge_fix}

We can explicitly show that all gauge fields are Stueckelberg ones at the level of BRST formulation.

Let us introduce an additional grading
\begin{equation}
	\deg{z_i} = 1\,, \qquad \deg{a^a_i} = \deg{b_i} = \deg{c_i} = 2\,.
\end{equation}
The BRST operator \eqref{b_BRST_massive} decomposes as $\brst_m = \brst_{-1} + \brst_{0} + \brst_1$ with
\begin{equation}
  \brst_{-1}
= m z_i \frac{\partial}{\partial b_i}\,,\quad
  \brst_0
= c_0 \left( \Box + m^2 \right)
+ c_i D^i
+ D^\dag_i \frac{\partial}{\partial b_i}
- c_i \frac{\partial}{\partial b_i} \frac{\partial}{\partial c_0}\,,\quad
  \brst_1
= m c_i \frac{\partial}{\partial z_i}\,.
\end{equation}

The reduced BRST operator is
\begin{equation}
	\widetilde \brst_m
= c_0 \left( \Box + m^2 \right)
+ c_i D^i
\end{equation}
acting on a subspace singled out by equations
\begin{equation}
  \widehat N_i \varphi = s_i \varphi\;, \qquad
  \widehat N_i{}^j \varphi = 0 \quad (i < j)\;, \qquad
  T^{ij} \varphi = 0\;,
\end{equation}
where
\begin{equation}
  \widehat N_i{}^j = N_i{}^j + c_i \frac{\partial}{\partial c_j}\;,\qquad
  \widehat N_i = \widehat N_i{}^i\,.
\end{equation}

See appendix \ref{app:Dimensional_reduction} for more details.

There are no negative ghost elements, so $H^0(\brst_m)$ is just kernel of $\brst_m$ restricted to the ghost-free subspace. I.e. the set of solutions of
\begin{equation} \label{b_Massive_max_covariant}
  \left( \Box + m^2 \right) \varphi^{(0)}
= D^i \varphi^{(0)}
= T^{ij} \varphi^{(0)}
= \left( N_i - s_i \right) \varphi^{(0)}
= N_k{}^l \varphi^{(0)} = 0 \quad (k < l)\,.
\end{equation}
Which is the maximal covariant set of equations describing a bosonic massive field of $(s_1, s_2, \ldots, s_n)$ symmetry.

In other words we showed that the gauge \eqref{b_Massive_max_covariant} in the gauge system described in the section \ref{sec:b_Dimensional_reduction} is reachable.

\subsubsection{Casimir operators revisited} \label{sec:b_massive_Casimir_indirect}

Let us compute the values of Casimir operators \eqref{Casimirs} acting on the space of gauge inequivalent field configurations defined in section \ref{sec:b_Dimensional_reduction} in other way. Let us denote the whole space as $V$, the subspace of pure gauges as $P$. $P$ is invariant under action of \eqref{Casimirs} in representation \eqref{regular_representation}, so Casimir operators act on $V / P$.

Note that by construction of BRST complex $V / P \simeq H^0(\brst_m) \simeq H^0(\widetilde \brst_m)$. 
Also note that we can consider $\cE$ as a subspace of ghost extended space $\widehat V$ where $\brst_m$ acts. So there are embeddings $V \hookrightarrow \cE \hookrightarrow \widehat V$.

There are no negative degree ghost variables in $\cE$. So $H^0(\widetilde \brst_m)$ is just the subspace of tensor fields $\phi$ singled out by \eqref{b_Massive_max_covariant}. By direct application of \eqref{Casimir2_regular} and \eqref{Casimir4_regular} we get
\begin{gather}
  C_2 \phi = - m^2 \phi\,,\\
	C_4 \phi =
- m^2 \sum\limits_{i = 1}^{n} s_i \left( s_i + d - 1 - 2 i \right) \phi\,.
\end{gather}


\section{Fermionic fields}

\subsection{Algebraic preliminaries}

\subsubsection{Spinor-tensor fields} \label{sec:f_Spinor-tensor}

We extend the algebra introduced in section \ref{sec:b_Algebraic_preliminaries} by Grassmann odd variables $\theta^a$ satisfying the canonical (anti)commutation relations
\begin{equation}
	\{ \fo^a, \fo^b \} = 2 \eta^{a b}\,.
\end{equation}

Consider the linear space $\cP^d_n(a) = \cS \otimes \bC[a_I]$, where $\cS$ is the Dirac representation of the Clifford algebra generated by $\fo^a$ and $\bC[a_I]$ is the space of polynomials in $a_I$.

The component decomposition of a general element $\psi(a) \in \cP^d_n(a)$ has the form
\begin{equation}
\begin{gathered}
  \psi(a) = \psi^\alpha(a) e_\alpha\,,\\
  \psi^\alpha(a) = \sum_{m_0 = 0}^{M_0} \ldots \sum_{m_n = 0}^{M_n} \psi^\alpha{}_{a_1\;\ldots\; a_{m_0};\;\ldots\ldots\;;\, b_1 \;\ldots\; b_{m_{n}}} a^{a_1}_0 \ldots a^{a_{m_0}}_0 
  \;\ldots\;
  a^{b_1}_{n} \ldots a^{b_{m_{n}}}_{n}\;,
\end{gathered}
\end{equation}
where $M_0, \ldots M_n \in \mathbb N$,
$e_\alpha$ is a basis in $\cS$ and $\alpha = 1, \ldots, 2^{[d/2]}$ is the Dirac spinor index.

The natural action of the associative algebra generated by $a^a_I$, $\bar a^J_b$ on $\cP^d_{n}(a_I)$ is defined by
\begin{equation} \label{f_Generators_action}
	a_I^a \psi(a) \coloneqq a_I^a \phi(a)\,,\qquad
  \bar a^I_a \phi(a) \coloneqq \frac{\partial}{\partial a_I^a} \phi(a)\,,\qquad
  \fo^a \psi^\alpha(a) \coloneqq (\gamma^a)^\alpha{}_\beta \psi^\beta(a)\,,
\end{equation}
where the gamma matrices $\gamma^a$ in basis $e_\alpha$ in $\cS$ are defined by $\fo^a e_\beta = (\gamma^a)^\alpha{}_\beta e_\alpha$.

\subsubsection{Lorentz algebra and orthosymplectic superalgebra}

We embed the Lorentz algebra $\Liealg{so}(d-1,1)$ as a Lie subalgebra into the above associative algebra by postulating its basis elements to be
\begin{equation}
  M_{ab} = a_I{}_a \bar a^{I}_b - a_I{}_b \bar a^{I}_a + \frac14 (\fo_a \fo_b - \fo_b \fo_a)\,.
\end{equation}
This also defines the representation of $\Liealg{so}(d-1,1)$ on $\cP^d_{n}(a_I)$ via \eqref{b_Generators_action}.

Simultaneously, the orthosymplectic superalgebra  $\Liealg{osp}(1|2n+2)$ can also be embedded into that associative algebra, and, hence, is also represented on $\cP^d_{n}(a_I)$. The basis elements are given by
\begin{equation} \label{f_osp_even}
	T_{IJ} = a_I^a a_{Ja}\,,\qquad
  T_I{}^J = \frac12 \left(a^a_I \bar a^J_a + \bar a^J_a a^a_I\right),\qquad
  T^{IJ} = \bar a^I_a \bar a^{Ja}\,,
\end{equation}
\begin{equation} \label{f_osp_odd}
	\Upsilon_I = a_I^a \fo_a\,, \qquad \Upsilon^I = \bar a^I_a \fo^a
\end{equation}
with the commutation relations given in Appendix \bref{app:comm}.
The two algebras mutually commute forming a reductive dual pair~\cite{Howe}.
So the space $\cP^d_{n}(a_I)$ is now $\Liealg{so}(d-1,1)-\Liealg{osp}(1|2n+2)$ bimodule.

\subsubsection{Poincar\'e algebra}

We realize the Poincar\'e algebra $\Liealg{iso}(d-1,1)$ on the same set of auxiliary variables. To this end, we split the original  variables as $a^a_0\equiv x^a,\;a^a_I\equiv a^a_i\,,\,I>0$ with $i=1,...,n$. Then, translations and Lorentz rotations are given by 
\begin{equation}
  P_a = \partial_a\,, \quad\qquad 
  M_{ab}
  = x_a \partial_b - x_b \partial_a
  + a_i{}_a \bar a^{i}_b - a_i{}_b \bar a^{i}_a
  + \frac14 (\fo_a \fo_b - \fo_b \fo_a)\,.
\end{equation}
They naturally act in the space $\cP_n^d(x,a)$ of smooth functions in $x^a$ with values in $\cP_n^d(a_i)$. 

Let us introduce special notation for the odd basis element of $\Liealg{osp}(1|2n+2)$ algebra
\begin{equation}
	\Dirac \equiv \Upsilon^0 = \fo^a \d_a
\end{equation}
in addition to \eqref{b_sp_basis_notation}.
From the $\Liealg{osp}(1|2n+2)$ graded commutation relations we have $\{\Dirac, \Dirac\} = 2 \Box$ meaning that the Dirac operator $\Dirac$ squares to the Klein-Gordon operator $\Box$.

\subsection{Massless fields}

The constraints are imposed on a spinor-tensor field $\psi \in \cP_n^d(x,a)$.

\paragraph{Differential constraint}
is just the Dirac equation
\begin{equation}
  \Dirac \psi = 0\,.
\end{equation}
\paragraph{Algebraic constraints}
now read as
\begin{equation}
  \Upsilon^i \psi = 0\,, \qquad
  N_i \psi = s_i \psi\,, \qquad
  N_i{}^j \psi = 0 \quad (i < j)\,.
\end{equation}
The spin weight conditions imposed on each type of  auxiliary variables constrain functions $\psi$ to be homogeneous polynomials in $a_i$. The Young symmetry and gamma-tracelessness conditions are the standard irreducibility  conditions for the $\Liealg{o}(d-1,1)$-representation of spin $s_1+\frac12 \geq s_2 + \frac12 \geq ... \geq s_n + \frac12$, where $s_i \in \mathbb{N}$.

Note that the Klein-Gordon operator, the divergence and the trace conditions are imposed by virtue of $\Dirac^2 = \Box$, $\;\{\Dirac, \Upsilon^i\} = 2 D^i$, $\;\{ \Upsilon^i, \Upsilon^j \} = 2 T^{ij}$.

\paragraph{Gauge transformations}
read as
\begin{equation}
  \delta \psi = D^\dag_i \chi^i\,,
\end{equation}
where the gauge parameters $\chi^i$ satisfy the same constraints as the fields $\psi$ except for the spin weight and Young symmetry constraints which are replaced by
\begin{gather}
\label{helcon2}
  N_i \chi^j = (s_i - \delta_i^j) \chi^j,\qquad
  N_i{}^j \chi^k = - \delta^k_i \chi^j \quad (i < j)\,.
\end{gather}

\subsubsection{Extended triplet formulation}

We impose all the differential constraints my means of the BRST operator, while the algebraic constraints, or more precisely, their BRST invariant extensions, we impose directly on the representation space.

We introduce anticommuting ghost variables $b_i, c_0, c_i$, $i = 1, ..., n$ and commuting ghost variable $\alpha$ with ghosts numbers $\gh{b_i} = -1$, $\gh{c_0} = \gh{c_i} = \gh{\alpha} = 1$. As $\alpha$ is a commuting variable there is an ambiguity in the functional class to work with. We choose functions $\Psi(x,a|\alpha,c_0,c,b)$ to be polynomials in $\alpha$.

The extended triplet BRST operator for fermionic helicity fields is given by
\begin{equation} \label{f_BRST}
  \brst
= \alpha \Dirac
+ c_0 \Box
+ c_i D^i
+ D^\dag_i \frac{\partial}{\partial b_i}
- \alpha \alpha \frac{\partial}{\partial c_0}
- c_i \frac{\partial}{\partial b_i} \frac{\partial}{\partial c_0}\,.
\end{equation}
It is defined on the subspace singled out by the following BRST-invariant extended constraints
\begin{equation} 
\label{f_alg}
  \widetilde N_i \Psi = s_i \Psi\,, \qquad
  \widetilde N_i{}^j \Psi = 0 \quad (i < j)\,, \qquad
  \widetilde \Upsilon^i \Psi = 0\,,
\end{equation}
where
\begin{equation}
  \widetilde N_i{}^j = N_i{}^j + b_i \frac{\partial}{\partial b_j} + c_i \frac{\partial}{\partial c_j}\,,\qquad
  \widetilde N_i = \widetilde N_i{}^i\,,\qquad
  \widetilde \Upsilon^i = \Upsilon^i - 2 \alpha \frac{\partial}{\partial c_i} + \frac{\partial}{\partial \alpha} \frac{\partial}{\partial b_i}\,.
\end{equation}
Note that the BRST operator \eqref{f_BRST} is well-defined and is nilpotent on the entire representation space and not only on the subspace \eqref{f_alg}.

\subsubsection{Triplet formulation} \label{sec:f_massless_Triplet}

Following similar steps as in section \ref{sec:b_Metric-like} we choose an additional grading to be the homogeneity degree in $c_0$.
The BRST operator \eqref{f_BRST} decomposes as
\begin{equation}
	\brst_{-1} = - \left( \alpha \alpha + c_i \frac{\partial}{\partial b_i} \right) \frac{\partial}{\partial c_0}\,,\qquad
  \brst_0 = \alpha \Dirac
+ c_i D^i
+ D^\dag_i \frac{\partial}{\partial b_i}\,,\qquad
  \brst_1 = c_0 \Box\,.
\end{equation}

The $\brst_{-1}$ cohomology analysis is surprisingly easier than in the bosonic case.

The cohomology of $\brst_{-1}$ is concentrated in degree $0$ and can be realized as a subspace $\cE$ of $c_0$-independent elements which are at most linear in $\alpha$. The space of fields with values in this subspace is equipped with the induced BRST operator $\widetilde \Omega$ which in this case is simply $\Omega_0$ defined on the equivalence classes. In terms of representatives which are at most linear in $\alpha$ it is given by 
\begin{equation}
  \widetilde\brst (\psi_0 + \alpha \psi_1)
= \alpha \Dirac \psi_0
- c_i\dl{b_i} \Dirac \psi_1
+ \left( c_i D^i + D^\dag_i \frac{\partial}{\partial b_i} \right) (\psi_0 + \alpha \psi_1)\,.
\end{equation}
The  second term  arises from $\alpha^2 \Dirac \psi_1$ which does not belong to $\cE$, and, hence, one needs to pick another representative of the same equivalence class.

Equivalently one can identify term $\psi_0 + \alpha \psi_1$ linear in $\alpha$ with a column $\begin{bmatrix} \psi_0 \\ \psi_1 \end{bmatrix}$ and operator $\widetilde \brst$ with a matrix
\begin{equation}
  \begin{bmatrix}
    c_i D^i + D^\dag_i \frac{\partial}{\partial b_i} & - c_i \frac{\partial}{\partial b_i} \Dirac \\
    \Dirac & c_i D^i + D^\dag_i \frac{\partial}{\partial b_i}
  \end{bmatrix}
\end{equation}
which acts on such columns with a matrix multiplication.


Let us perform component analysis. Elements of $\cE$ of ghost numbers $0$ and $-1$ are given respectively by
\begin{multline}
  \Phi = \psi + \sum_{k=1}^{n} c_{i_1} \ldots c_{i_{k}} b_{j_1} \ldots b_{j_k} \lambda^{i_1 \ldots i_{k} | j_1 \ldots j_k} + \\
  + \alpha \sum_{k=1}^{n} c_{i_1} \ldots c_{i_{k-1}} b_{j_1} \ldots b_{j_k} \chi^{i_1 \ldots i_{k-1} | j_1 \ldots j_k}\,,
\end{multline}
\begin{multline}
\Xi = b_j \epsilon^j + \sum_{k=2}^{n} c_{i_1} \ldots c_{i_{k-1}} b_{j_1} \ldots b_{j_k} \epsilon^{i_1 \ldots i_{k-1} | j_1 \ldots j_k} + \\
+ \alpha \sum_{k=2}^{n} c_{i_1} \ldots c_{i_{k-2}} b_{j_1} \ldots b_{j_k} \xi^{i_1 \ldots i_{k-2} | j_1 \ldots j_k}\;.
\end{multline}
From the algebraic constraints \eqref{f_alg} it follows that the lowest components $\psi$ and $\epsilon^i$ satisfy the triple trace conditions 
\begin{equation} \label{f_triple_trace}
  \Upsilon^{(i} \Upsilon^{j} \Upsilon^{k)} \psi = 0\,,\qquad
  \Upsilon^{(i} \epsilon^{j)} = 0\,,
\end{equation}
as well as the Young symmetry and spin conditions
\begin{equation} \label{f_Young_and_spin}
  N_i{}^j \psi = 0 \;\;  (i<j)\,,
  \;\;\;
  N_i \psi = s_i \psi\,,
  \;\;\;
  N_i{}^j \epsilon^k = - \delta^k_i \epsilon^j \;\; (i < j)\,,
  \;\;\;
  N_i \epsilon^j = (s_i - \delta_i^j) \epsilon^j\,.
\end{equation}
One concludes that $\psi$ and $\epsilon^i$ are precisely the original Fang-Fronsdal-Labastida spinor-tensor fields and their associated gauge parameters \cite{Fang:1978wz,Labastida:1986zb,Labastida:1989kw}.

The consequence of $\widetilde \brst \Psi = 0$ is
\begin{equation}
  \Dirac \psi + D^\dag_i \chi^{|i} = 0\,.
\end{equation}
The BRST-extended gamma-trace conditions \eqref{f_alg} imply $\chi^{|i} = - \Upsilon^i \psi$ thereby giving the reduced equations of motion
\begin{equation} \label{f_metric-like}
  \left( \Dirac - D^\dag_i \Upsilon^i \right) \psi= 0\,.
\end{equation}
This is the Fang-Fronsdal-Labastida equations for massless mixed-symmetry fermionic fields \cite{Fang:1978wz,Labastida:1986zb,Campoleoni:2009gs}. Note that, just like the standard Dirac equation,  the Fang-Fronsdal-Labastida equation can be squared, resulting in
\begin{equation}
  \left( \Box - D^\dag_i D^i + \frac12 D^\dag_i D^\dag_j T^{ij} \right) \psi = 0\,,
\end{equation}
which is the Labastida equations for massless mixed-symmetry bosonic fields \cite{Labastida:1989kw}. Here we made use 
of $\Dirac D^\dag_i \Upsilon^i \psi = \Box \psi$ which is the result of acting by $\Dirac$ on \eqref{f_metric-like}.

By construction, the reduced equations \eqref{f_metric-like} are invariant with respect to the gauge transformations
\begin{equation}
  \delta \psi = D^\dag_i \epsilon^i\,,
\end{equation}
where the gauge fields and parameters satisfy the algebraic conditions \eqref{f_triple_trace}, \eqref{f_Young_and_spin}.

\subsubsection{Triplet Lagrangian} \label{sec:f_massless_Lagrangian}

Tensoring the Fock space from section \ref{sec:b_Lagrangian} with $\cS$ (see section \ref{sec:f_Spinor-tensor}) and equipping $\cS$ with an inner product such that $(\theta^a)^\dagger = -\theta^a$ we end up with the space equipped with an inner product $\inner{}{}^\prime$. Finally, the formal inner product on $\cE$ is taken to be
\begin{equation} \label{f_inner}
	\langle \phi, \psi \rangle \equiv \langle \phi_0 + \alpha \phi_1, \psi_0 + \alpha \psi_1 \rangle
  \coloneqq
  \int \dif^d x \big( \langle \phi_0, \psi_1 \rangle^\prime + \langle \phi_1, \psi_0 \rangle^\prime \big)\,.
\end{equation}
It is straightforward to see that the inner product \eqref{f_inner} in nondegenerate and $\widetilde \Omega$ is formally symmetric with respect to it.

The equations of motion $\widetilde \brst \Psi^{(0)} = 0$ follow from the action
\begin{equation}
  S = \frac12 \langle \Psi^{(0)}, \widetilde \brst \Psi^{(0)} \rangle\,,
\end{equation}
which is invariant under gauge transformations $\delta \Psi^{(0)} = \widetilde \brst \Psi^{(-1)}$.
Supplemented with the off-shell algebraic constraints \eqref{f_alg} it is equivalent to the Fang-Fronsdal-Labastida action.

\subsection{Massive fields}

We again start from $(d+1)$-dimensional system describing arbitrary fermionic massless field with fixed momentum along the $(d+1)$-th direction: $(p^a, m)$.
Let us also introduce notations for odd and even oscillators ''along that direction'': $(\fo^a, \ao)$, $(a^a_i, z_i)$.

\paragraph{Differential constraint}
We still have the Dirac equation 
\begin{equation} \label{f_massive_EOM_diff}
  \left( \Dirac + m \ao \right) \psi = 0\;.
\end{equation}
\paragraph{Algebraic constraints}
become
\begin{equation} \label{f_massive_EOM_alg}
  \left( \Upsilon^i + \ao \frac{\partial}{\partial z_i} \right) \psi = 0\,,
  \quad
  \left( N_i + Z_i \right) \psi = s_i \psi\,,
  \quad
  \left( N_i{}^j + z_i \frac{\partial}{\partial z_j}\right) \psi = 0 \quad (i < j)\,.
\end{equation}

\paragraph{Gauge transformaions}
read
\begin{equation} \label{f_massive_gauge}
  \delta \psi = \left( D^\dag_i + m z_i \right) \chi^i\,,
\end{equation}
where the gauge parameters $\chi^i$ satisfy the same constraints as the fields $\psi$ except for the spin weight and Young symmetry constraints which are replaced by
\begin{gather} \label{f_gauge_params_algebraic_constraints}
  \left( N_i + Z_i \right) \chi^j = (s_i - \delta_i^j) \chi^j,\qquad
  \left( N_i{}^j + z_i \frac{\partial}{\partial z_j} \right) \chi^k = - \delta^k_i \chi^j \quad (i < j)\,.
\end{gather}

Note the straightforward consequences of the constraints: $(\Dirac + m \ao)^2 = \Box + m^2$, $\{\Dirac + m \ao, \Upsilon^i + \ao \frac{\partial}{\partial z_i}\} = 2 \left( D^i + m \frac{\partial}{\partial z_i} \right)$, $\{ \Upsilon^i + \ao \frac{\partial}{\partial z_i}, \Upsilon^j + \ao \frac{\partial}{\partial z_j} \} = 2 \left( T^{ij} + \frac{\partial}{\partial z_i} \frac{\partial}{\partial z_j} \right)$.

Also note that one can get rid of $\ao$ by automorphism of the Clifford algebra $\theta \mapsto \theta \ao$, restoring ''classical'' form of the Dirac equation.

There is a slight difference in what $\ao$ is, depending on whether the spacetime dimension $d$ is even or odd. More precisely, for even $d$ the $\ao$ is just the "fifth gamma" 
\begin{equation}
  \ao
= \frac{i^{d/2 - 1}}{d!} \sqrt{- \det \eta_{ab}}\, \varepsilon_{a_1 \ldots a_d} \fo^{a_1} \ldots \fo^{a_d}
= i^{d/2 - 1} \fo^0 \fo^1 \ldots \fo^{d - 1}\,,
\end{equation}
that is $\ao$ is realized in terms of the reduced Clifford algebra generated by $\fo^0, \ldots, \fo^{d-1}$ and its module. In odd $d$ it is not the case and $\ao$ extends the reduced Clifford algebra to $\{\fo^A, \fo^B\} = 2 \eta^{AB}$, where $A = (a, d)$, and $\eta^{dd} = 1$, and $\fo^{d} \coloneqq \ao$. In this case, the $d$-dimensional spinor representation splits to the right and left spinor spaces, and, hence, the spectrum of fields is duplicated. However, the original $(d+1)$-dimensional Clifford algebra is even dimensional, and, therefore, there is a new "fifth gamma" $\widetilde \ao$ that can be used to project out a half of the spinor components via the standard $P_{\pm} = \half(1 \pm \widetilde\ao)$. For simplicity, throughout the paper we explicitly treat the case of even $d$ unless otherwise indicated.

\subsubsection{Casimir invariant}

Analogously to the bosonic case we use the explicit expressions \eqref{Casimir2_regular} and \eqref{Casimir4_regular} for the quadratic and quartic Casimir operators in terms of the $\Liealg{osp}(1|2n)$ basis elements and take into account constraints \eqref{f_massive_EOM_diff}, \eqref{f_massive_EOM_alg}. As \eqref{Casimir4_regular} splits into sum of bosonic and fermionic part, we have
\begin{equation}
	C_2 \psi = - m^2 \psi\,,
\end{equation}
\begin{equation}
	C_4 \psi = B + F\,,
\end{equation}
where $B$ is the same as in the RHS of \eqref{b_massive_C4} and
\begin{equation}
	F = - m^2 \left( \sum_i s_i + \frac{(d - 1) (d - 2)}{8} \right) \psi + m \left( D^\dag_i + m z_i \right) \frac{\partial}{\partial z_i} \psi\,.
\end{equation}
The last term is pure gauge.

So the action of $C_4$ is well defined and diagonal on the space of equivalence classes $\overline \psi$ of field configurations modulo gauge transformations:
\begin{equation}
	C_4 \overline \psi =
- m^2 \left( \sum_{i = 1}^n s_i (s_i + d - 2 i) + \frac{(d - 1) (d - 2)}{8} \right) \overline \psi\,.
\end{equation}

\subsubsection{Extended triplet formulation}

The BRST operator associated to the constrained system \eqref{f_massive_EOM_diff}, \eqref{f_massive_EOM_alg}, \eqref{f_massive_gauge}, is
\begin{equation} \label{f_BRST_massive}
  \brst_m
= \alpha \left( \Dirac + m \ao \right)
+ c_0 \left( \Box + m^2 \right)
+ c_i \left( D^i + m \frac{\partial}{\partial z_i} \right)
+ \left( D^\dag_i + m z_i \right) \frac{\partial}{\partial b_i}
- \alpha \alpha \frac{\partial}{\partial c_0}
- c_i \frac{\partial}{\partial b_i} \frac{\partial}{\partial c_0}\,.
\end{equation}
It is a massive analogue of \eqref{f_BRST}.

It acts on a subspace of $\cP_n^d(x, a, z) \otimes \mathbb R[b_i,c_0,c_i] \ni \Psi(x, a \,|\, b_i, c_0, c_i)$ singled out by the BRST invariant extensions of algebraic constraints \eqref{f_massive_EOM_alg}
\begin{equation} \label{f_massive_Triplet_algebraic}
\begin{aligned}
	\left( N_i{}^j + z_i \frac{\partial}{\partial z_j} + b_i \frac{\partial}{\partial b_j} + c_i \frac{\partial}{\partial c_j} \right) \Psi &= 0 \quad (i < j)\,,\\
  \left( N_i + Z_i + b_i \frac{\partial}{\partial b_i} + c_i \frac{\partial}{\partial c_i} \right) \Psi &= s_i \Psi\,,\\
  \left( \Upsilon^i + \ao \frac{\partial}{\partial z_i} - 2 \alpha \frac{\partial}{\partial c_i} + \frac{\partial}{\partial \alpha} \frac{\partial}{\partial b_i} \right) \Psi &= 0\,.
\end{aligned}
\end{equation}

\subsubsection{Triplet formulation}

Following similar steps as in section \ref{sec:f_massless_Triplet} we choose an additional grading to be the homogeneity degree in $c_0$.
The BRST operator \eqref{f_BRST_massive} decomposes as
\begin{equation}
\begin{gathered}
	\brst_{-1} = - \left( \alpha \alpha + c_i \frac{\partial}{\partial b_i} \right) \frac{\partial}{\partial c_0}\,,\\
  \brst_0 = \alpha \left( \Dirac + m \ao \right)
+ c_i \left( D^i + m \frac{\partial}{\partial z_i} \right)
+ \left( D^\dag_i + m z_i \right) \frac{\partial}{\partial b_i}\,,\\
  \brst_1 = c_0 \left( \Box + m^2 \right).
\end{gathered}
\end{equation}
So the reduced BRST operator acts on the space $\cE$ which is identified with $c_0$-free terms which are at most linear in $\alpha$ is
\begin{multline}
 \widetilde \brst_m (\psi_0 + \alpha \psi_1)
= \alpha \left( \Dirac + m \ao \right) \psi_0
- c_i \frac{\partial}{\partial b_i} \left( \Dirac + m \ao \right) \psi_1\\
+ \left(
    c_i \big( D^i + m \frac{\partial}{\partial z_i} \big)
  + \big( D^\dag_i + m z_i \big) \frac{\partial}{\partial b_i}
  \right)
  (\psi_0 + \alpha \psi_1)\,.
\end{multline}

\subsubsection{Lagrangian formulation}

The equations of motion $\widetilde \brst_m \Psi^{(0)} = 0$ follows from the action
\begin{equation}
  S = \frac12 \langle \Psi^{(0)}, \widetilde \brst_m \Psi^{(0)} \rangle\,,
\end{equation}
which is invariant under gauge transformations $\delta \Psi^{(0)} = \widetilde \brst_m \Psi^{(-1)}$.

The inner product it essentially the same as in section \ref{sec:f_massless_Lagrangian} with conjugation rules for additional oscillators
\begin{equation}
	(z_i)^\dag = - \frac{\partial}{\partial z_i}\,,\qquad
	\ao^\dag = - \ao\,.
\end{equation}

\subsubsection{Gauge fixing}

We can explicitly show that all gauge fields are Shtueckelberg ones at the level of BRST formulation following the same lines as in section \ref{sec:b_Gauge_fix}.

We introduce an additional grading
\begin{equation}
	\deg{z_i} = 1\,, \qquad \deg{a^a_i} = \deg{b_i} = \deg{c_i} = 2\,.
\end{equation}
The BRST operator \eqref{f_BRST_massive} decomposes as $\brst_m = \brst_{-1} + \brst_{0} + \brst_1$ with
\begin{equation}
\begin{gathered}
  \brst_{-1}
= m z_i \frac{\partial}{\partial b_i}\,,\\
  \brst_{0}
= \alpha \left( \Dirac + m \ao \right)
+ c_0 \left( \Box + m^2 \right)
+ c_i D^i
+ D^\dag_i \frac{\partial}{\partial b_i}
- \alpha \alpha \frac{\partial}{\partial c_0}
- c_i \frac{\partial}{\partial b_i} \frac{\partial}{\partial c_0}\,,\\
  \brst_1
= m c_i \frac{\partial}{\partial z_i}\,.
\end{gathered}
\end{equation}
Analoguosly to the bosonic case $z_i$ and $b_i$ go away in pairs, so the reduced BRST operator is
\begin{equation}
	\widetilde \brst_m
= \alpha \left( \Dirac + m \ao \right)
+ c_0 \left( \Box + m^2 \right)
+ c_i D^i
- \alpha \alpha \frac{\partial}{\partial c_0}
\end{equation}
which acts on a subspace singled out by equations
\begin{equation}
  \widehat N_i \psi = s_i \psi\;, \qquad
  \widehat N_i{}^j \psi = 0 \quad (i < j)\;, \qquad
  \widehat \Upsilon^i \psi = 0\;,
\end{equation}
where
\begin{equation}
	\widehat N_i{}^i = N_i{}^j  + c_i \frac{\partial}{\partial c_j} \quad (i < j)\,,\qquad
  \widehat N_i = \widehat N_i{}^i\,,\qquad
  \widehat \Upsilon^i = \Upsilon^i - 2 \alpha \frac{\partial}{\partial c_i}\,.
\end{equation}

There are no negative ghost degree elements in that subspace, so $H^0(\widetilde \brst_m)$ is just kernel of $\widetilde \brst_m$ restricted to the ghost-free subspace. I.e. the set of solutions of
\begin{equation} \label{f_Massive_max_covariant}
	( \Dirac + m \ao ) \psi^{(0)}
= \Upsilon^i \psi^{(0)}
= \left( N_i - s_i \right) \psi^{(0)}
= N_i{}^j \psi^{(0)} = 0 \quad (i < j)
\end{equation}
($(\Box + m^2) \psi^{(0)} = D^i \psi^{(0)} = 0$ are consequences).
This is the maximal covariant set of equations describing a fermionic massive field of $s_1 + \frac12 \ge s_2 + \frac12 \ge \ldots \ge s_n + \frac12$ helicity.

\section{Conclusion}

We build a systematic description of both bosonic and fermionic massive fields of arbitrary symmetry via dimensional reduction. We used BRST language, which allowed us easily move between different equivalent formulations. In particular this was important in the fermionic case where it helped to arrive to additional formulation which possess a natural inner product, and leaded straightforwardly to Lagrangian.

\vspace{5mm}
\noindent \textbf{Acknowledgements.} I am grateful to K. Alkalaev and M. Grigoriev for suggesting the problem and useful discussions. The work was supported by the grant RFBR No 18-02-01024.

\appendix

\section{$\Liealg{osp}(1|2n)$ commutation relations} \label{app:comm}

The basis $\Liealg{osp}(1|2n)$ elements are defined in \eqref{f_osp_even} and \eqref{f_osp_odd}. Their non-zero commutation relations  in the even sector are
\be
\label{Brela}
\ba{c}
\dps
[T_I{}^J, T_K{}^L]= \delta_K^J T_I{}^L-\delta_I^LT_K{}^J\,,
\quad
[T^{IJ}, T_{KL}] = \delta^I_K T_L{}^J+\delta^I_L T_K{}^J+\delta^J_K T_L{}^I+\delta^J_L T_K{}^I\,,
\\
\\
\dps
[T_K{}^L, T_{IJ}]= \delta_J^L T_{KI}+\delta^L_IT_{KJ} \;,
\quad
[T^{IJ}, T_K{}^L]=\delta^I_K T^{JL}+ \delta^J_KT^{IL} \;,
\ea
\ee
in the odd sector are
\begin{equation} \label{Frela}
  \{\Upsilon_I, \Upsilon_J\} = 2\, T_{IJ}\;,
  \qquad
  \{\Upsilon_I, \Upsilon^J\} = 2\, T_{I}{}^{J}\;,
  \qquad
  \{\Upsilon^I, \Upsilon^J\} = 2\, T^{IJ}\;,
\end{equation}
in the cross-sector are
\begin{equation} \label{Crela}
\begin{array}{c}
  [T_{IJ}, \Upsilon^K] = -\delta_I^K \Upsilon_J  - \delta_J^K \Upsilon_K\;,
  \qquad
  [T^{IJ}, \Upsilon^K] = 0\;,
  \qquad
  [T_I{}^J, \Upsilon^K] = -\delta_I^K \Upsilon_J\;,
  \\
  \\
  
  [T^{IJ}, \Upsilon_K] = \delta_I^K \Upsilon^J + \delta_K^J \Upsilon^I\;,
  \qquad
  [T_{IJ}, \Upsilon_K] = 0\;,
  \qquad
  [T_I{}^J, \Upsilon_K] = \delta_K^J \Upsilon_I\;.
\end{array}
\end{equation}

\section{Casimir operators} \label{app:Casimir}

The quadratic and quartic Casimir operators of the $\Liealg{iso}(p, q)$ algebra are
\begin{equation} \label{Casimirs}
C_2 \Big( \mathfrak{iso}(p, q) \Big) = P_a P^a \equiv P^2\,,
\qquad
C_4 \Big( \mathfrak{iso}(p, q) \Big) = M_{ab} P^b M^{ac} P_c - \frac12 M^2 P^2\,,
\end{equation}
where $P_a$ stands for translation and $M_{ab}$ for rotation generators, respectively. In what follows we express \eqref{Casimirs} in terms of $\Liealg{osp}$ basis elements.

\paragraph{Regular spinor-tensor representation.}
Let $\Liealg{iso}(d-1,1)$ basis elements $P_a\,, M_{ab}\,, a,b = 0, ..., d-1$ act on \{spinor\}-tensor fields as
\begin{equation} \label{regular_representation}
  P_a = \partial_a\,, \quad\qquad 
  M_{ab}
  = x_a \partial_b - x_b \partial_a
  + a_i{}_a \bar a^{i}_b - a_i{}_b \bar a^{i}_a
  + \left\{ \frac14 (\fo_a \fo_b - \fo_b \fo_a) \right\}\,,
\end{equation}
$i = 1, \ldots, n$. Expressing quadratic and quartic Casimir operators in terms of the $\Liealg{osp}(1|2n+2)$ basis elements we find 
\begin{equation} \label{Casimir2_regular}
  C_2 \Big( \Liealg{iso}(d-1,1) \Big) = \Box\,,
\end{equation}
\begin{multline} \label{Casimir4_regular}
  C_4 \Big( \Liealg{iso}(d-1,1) \Big)
  =
    \left( (d - n - 2) N_i{}^i + N_j{}^i N_i{}^j - T_{i j} T^{i j} \right) \Box\\
  + T_{i j} D^i D^j + (2 - d) D^\dag_i D^i - 2 D^\dag_j N_i{}^j D^i + D^\dag_i D^\dag_j T^{i j}\\
+ \left\{
    \left( \Upsilon_i D^i - D^\dag_i \Upsilon^i \right) \Dirac
  + \left( N_i{}^i - \Upsilon_i \Upsilon^i + \frac{(d - 1) (d - 2)}{8} \right) \Box
  \right\}\,.
\end{multline}
Dropping terms in curly brackets we get the bosonic Casimir operator. Also, the above $\Liealg{osp}(1|2n+2)$ representation holds for any $\Liealg{iso}(k,l)$ with $k+l = d$.

\section{Details of dimensional reduction} \label{app:Dimensional_reduction}

\subsection{Symmetric fields}

Let us start with the case of symmetric fields ($n = 1$).

\begin{equation} 
\label{helicity_massive_BRST_full_symmetric}
  \brst
= \alpha \left( \Dirac + m \ao \right)
+ c_0 \left( \Box + m^2 \right)
+ c \left( D + m \frac{\partial}{\partial z} \right)
+ \left( D^\dag + m z \right) \frac{\partial}{\partial b}
- \alpha \alpha \frac{\partial}{\partial c_0}
- c \frac{\partial}{\partial b} \frac{\partial}{\partial c_0}\,.
\end{equation}
The triplet BRST operator \eqref{helicity_massive_BRST_full_symmetric} acts on the subspace singled out by the BRST extended constraints
\begin{equation} 
\label{helicity_massive_algebraic_constraints_BRST_extended_symmetric}
  \widehat N_a \Psi = s \Psi\;, \qquad
  \widehat \Upsilon \Psi = 0\;,
\end{equation}
where
\begin{equation}
  \widehat N_a = N_a + z \frac{\partial}{\partial z} + b \frac{\partial}{\partial b} + c \frac{\partial}{\partial c}\;,\quad
  \widehat \Upsilon = \Upsilon + \ao \frac{\partial}{\partial z} - 2 \alpha \frac{\partial}{\partial c} + \frac{\partial}{\partial \alpha} \frac{\partial}{\partial b}\;.
\end{equation}

Let us introduce an additional grading
\begin{equation}
	\deg{z} = 1\,, \qquad \deg{a^a} = \deg{b} = \deg{c} = 2\,.
\end{equation}
The BRST operator \eqref{helicity_massive_BRST_full_symmetric} decomposes as $\brst = \brst_{-1} + \brst_{0} + \brst_1$ with
\begin{equation}
\begin{gathered}
  \brst_{-1} = m z \frac{\partial}{\partial b}\,,\\
  \brst_0
= \alpha \left( \Dirac + m \ao \right)
+ c_0 \left( \Box + m^2 \right)
+ c D
+ D^\dag \frac{\partial}{\partial b}
- \alpha \alpha \frac{\partial}{\partial c_0}
- c \frac{\partial}{\partial b} \frac{\partial}{\partial c_0} \,,\\
  \brst_1 = m c \frac{\partial}{\partial z}\,.
\end{gathered}
\end{equation}


The whole space can be decomposed as $\cE \oplus \cG \oplus \cF$ where $\cG = \Ima \Omega_{-1}$ and $\cE \oplus \cG = \Ker \Omega_{-1}$.
\begin{equation}
	\Psi
= E + G + F
= \varphi_0 + \sum_{k = 1}^{\infty} z^k \varphi_k + b \sum_{k = 0}^{\infty} z^k \psi_k\,.
\end{equation}
\begin{equation}
  \overset{\cG \cF}{\Omega} \Psi \equiv (\Omega F)|_\cG
= \sum_{k = 1}^{\infty} z^k \left( \left( D^\dag + c \frac{\partial}{\partial c_0} \right) \psi_k + m \psi_{k-1} \right)
\end{equation}
is algebraically invertible.
\begin{equation}
	\overset{\cE \cE}{\Omega} \Psi \equiv (\Omega E)|_\cE
= \left( \alpha \left( \Dirac + m \ao \right) + c_0 \left( \Box + m^2 \right) + c D - \alpha \alpha \frac{\partial}{\partial c_0} \right) \varphi_0\,.
\end{equation}
\begin{equation}
	\overset{\cG \cE}{\Omega} \Psi \equiv (\Omega E)|_\cG = 0\,.
\end{equation}
So according to the general prescription the reduced BRST operator $\widetilde \Omega$ acting on a $H(\Omega_{-1}) \cong \cE$ is
\begin{equation}
	\widetilde \Omega = \overset{\cE \cE}{\Omega} - \overset{\cE \cF}{\Omega} (\overset{\cG \cF}{\Omega})^{-1} \overset{\cG \cE}{\Omega}
= \alpha \left( \Dirac + m \ao \right) + c_0 \left( \Box + m^2 \right) + c D - \alpha \alpha \frac{\partial}{\partial c_0}\,.
\end{equation}
It acts on the intersection of $\cE$ and the subspace selected by \eqref{helicity_massive_algebraic_constraints_BRST_extended_symmetric}. I.e.
\begin{equation}
	\left( N_a + c \frac{\partial}{\partial c} \right) \varphi = s \varphi\,, \qquad
  \left( \Upsilon - 2 \alpha \frac{\partial}{\partial c} \right) \varphi = 0\,.
\end{equation}
Note that this space does not have elements with negative ghost degree --- the gauge is fixed. And the component with vanishing ghost degree satisfies
\begin{equation}
	\left( \Dirac + m \ao \right) \varphi^{(0)} = 0\,,\qquad
  N_a \varphi^{(0)} = s \varphi^{(0)}\,,\qquad
  \Upsilon \varphi^{(0)} = 0\,.
\end{equation}
($(\Box + m^2) \varphi^{(0)} = D \varphi^{(0)} = 0$ are their consequences.)

\subsection{Mixed-symmetry fields}

One can consequently repeat argument from symmetric fields. Namely, to introduce grading
\begin{equation}
	\deg{z_i} = 1\,, \qquad \deg{a^a} = \deg{b_i} = \deg{c_i} = 2\,.
\end{equation}
for $i = 1\,,\ldots\,,n$ and consequently perform homological reduction for any $i$, effectively throwing out pairs of $z_i$ and $b_i$.

The result is the reduced BRST operator
\begin{equation}
	\widetilde \brst
= \alpha \left( \Dirac + m \ao \right)
+ c_0 \left( \Box + m^2 \right)
+ c_i D^i
- \alpha \alpha \frac{\partial}{\partial c_0}
\end{equation}
acting on a subspace
\begin{equation} 
  \widehat N_i \psi = s_i \psi\;, \qquad
  \widehat N_i{}^j \psi = 0 \quad (i < j)\;, \qquad
  \widehat \Upsilon^i \psi = 0\;,
\end{equation}
where
\begin{equation}
  \widehat N_i{}^j = N_i{}^j + c_i \frac{\partial}{\partial c_j}\;,\quad
  \widehat N_i = \widehat N_i{}^i\; \text{(no sum)},\quad
  \widehat \Upsilon^i = \Upsilon^i - 2 \alpha \frac{\partial}{\partial c_i}\;.
\end{equation}

The physical content is
\begin{equation}
	\left( \Dirac + m \ao \right) \psi^{(0)} = 0,\quad
  N_i \psi^{(0)} = s_i \psi^{(0)},\quad
  N_i{}^j \psi^{(0)} = 0 \quad (i < j),\quad
  \Upsilon^i \psi^{(0)} = 0.
\end{equation}
($(\Box + m^2) \psi^{(0)} = D^i \psi^{(0)} = 0$ are consequences.)



\providecommand{\href}[2]{#2}\begingroup\raggedright\endgroup

\end{document}